\documentclass[aps,prd,amsmath,superscriptaddress,twocolumn,floatfix,nofootinbib,preprintnumbers,10pt]{revtex4}
\usepackage{amssymb}\usepackage{txfonts}\usepackage{epsfig}\usepackage{bm}
\usepackage{color}\usepackage{graphicx,graphics}\usepackage{multirow}\usepackage{float}\usepackage{ulem}
\usepackage{setspace}\usepackage{slashed}\usepackage{booktabs}\usepackage{hyperref}
\hypersetup{colorlinks=true, citecolor=blue, linkcolor=red, filecolor=black,urlcolor=blue}
\allowdisplaybreaks[4]

\begin{document}

%\title{Study of fragmentation functions in the fully inclusive jet production electron-positron annihilation}
%\title{Probing Twist-3 Fragmentation Functions via fully inclusive jet Production \\in Electron-Positron Annihilation}
%\title{Probing twist-3 effects via fully inclusive jet production in $e^+e^-$ annihilation}
\title{Twist-3 effects in fully inclusive jet production from $e^+e^-$ annihilation}

\author{Weihua Yang}%\thanks{Contact author: yangwh@ytu.edu.cn}

\affiliation{College of Nuclear Equipment and Nuclear Engineering, Yantai University,\\ Yantai, Shandong 264005, China}

\begin{abstract}
Electron-positron annihilation is an ideal place to study the hadronization process which is described by fragmentation functions. In contrast to conventional approaches based on single-inclusive or semi-inclusive annihilation, the use of fully inclusive jets to study fragmentation functions is more advantageous. This approach provides novel insight into the underlying hadronization mechanisms. For example, the mass of the jet, which breaks chiral symmetry, can be effectively used to probe twist-3 chiral-odd effects. In this paper, we present a systematic theoretical framework for calculating the differential cross-section up to twist-3 for the unpolarized annihilation process. Calculations take into account both the electromagnetic interaction and weak interaction. We find that leading-twist fragmentation functions are convoluted into the twist-3 effects. This probably provides a new perspective for studying fragmentation functions. To this end,  we introduce two twist-3 azimuthal asymmetries and give the corresponding numerical estimates. We notice that the twist-3 effects are not significantly suppressed in high-energy regions due to the presence of the weak interaction. In other words, it is feasible to measure twist-3 effects at future high-energy electron-positron colliders to study hadronization mechanisms. 
\end{abstract}

\maketitle

\section{Introduction}\label{sec:introduction}

Quantum chromodynamics (QCD) is the fundamental non-Abelian gauge theory of the strong interaction, with a Lagrangian describing the dynamics of quarks, gluons, and their color-mediated interactions. It is defined by two features: asymptotic freedom and color confinement. Due to the latter, isolated quarks and gluons cannot be experimentally observed; instead, only color-singlet hadrons are detectable. The physical process by which partons transition into these color-singlet states is known as hadronization. Thus, studying hadronization mechanisms is of great significance for understanding QCD. Hadronization is inherently non-perturbative and cannot be calculated directly from first principles. Consequently, it is typically described through phenomenological models, such as the Feynman–Field model, LUND model, Webber model, and recombination models \cite{Field:1977fa,Andersson:1983ia,Marchesini:1983bm,Webber:1983if,Anisovich:1972pq,Bjorken:1973mh,Das:1977cp,Xie:1988wi}. To gain deeper insights, one often analyzes the behavior of single or paired hadrons produced during the process, which are encoded in hadron fragmentation functions (FFs) \cite{Berman:1971xz,Collins:1981uw} and di-hadron fragmentation functions (diFFs) \cite{Bianconi:1999cd,Bianconi:1999uc,Bacchetta:2003vn}, respectively. A comprehensive review of these functions is provided in Ref. \cite{Metz:2016swz}.
Compared to model descriptions, FFs play a central role in understanding hadronization, as they encode the probability for a parton (quark or gluon) to fragment into a specific hadron, providing critical information about the non-perturbative regime of QCD. Generally speaking, FFs can be studied in various high-energy processes, including semi-inclusive deep inelastic scattering (SIDIS), proton-proton collisions, and electron–positron annihilation. Unlike SIDIS and proton-proton collisions, where parton distribution functions (PDFs) also enter the measurements, electron–positron annihilation involves only FFs, providing a particularly clean and favorable environment for studying FFs and hadronization mechanisms.

Beyond the study of individual or paired hadrons, it is essential to investigate the collective behavior of multiple hadrons during hadronization. A jet is defined as an ensemble of hadrons clustered within a specific phase space characterized by localized parameters. In recent years, jets originating from quarks, anti-quarks, or gluons have emerged as powerful probes for understanding the hadronization mechanism \cite{Song:2010pf,Song:2013sja,Wei:2016far,Yang:2017sxz,Chen:2020ugq,Yang:2020qsk,Yang:2022knp, Yang:2022xwy,Yang:2023vyv,Yang:2023zod,Yang:2025hky, Gutierrez-Reyes:2018qez,Gutierrez-Reyes:2019vbx,Liu:2018trl,Liu:2020dct,Kang:2020fka,Arratia:2020ssx,Arratia:2019vju,Arratia:2022oxd}. This research has been further advanced by the development of jet fragmentation functions (JFFs) \cite{Procura:2009vm,Jain:2011xz,Jain:2011iu,Chien:2015ctp,Arleo:2013tya,Kaufmann:2015hma,Kang:2016ehg,Dai:2016hzf,Kang:2019ahe,Kang:2017glf,Kang:2020xyq}, which provide valuable insights into the internal hadron distribution within a jet. When considering jet production in high-energy reactions, a more comprehensive perspective is gained by focusing on fully inclusive jets \cite{Collins:2007ph,Accardi:2008ne,Accardi:2017pmi,Accardi:2020iqn,Accardi:2023cmh}. The introduction of fully inclusive jets significantly simplifies theoretical calculations, as only the kinematic properties of the final-state jet need to be reconstructed, rather than the individual hadronic final states. This approach not only provides a more direct probe of PDFs and FFs  but also allows for the exploration of  twist-3 effects. Moreover, the mass of the jet, which was usually neglected in previous studies \cite{Song:2010pf,Song:2013sja,Wei:2016far,Yang:2017sxz,Chen:2020ugq,Yang:2020qsk,Yang:2022knp, Yang:2022xwy,Yang:2023vyv,Yang:2023zod,Yang:2025hky}, offers a unique opportunity to study chiral-odd functions \cite{Accardi:2017pmi}, which are challenging to access in other types of hadronic processes.  
Given these considerations, in this paper we present a systematic theoretical framework for calculating the differential cross-section of fully inclusive jet production in the unpolarized electron–positron annihilation. The calculations are performed up to twist-3 accuracy, incorporating both electromagnetic and weak interactions. From the differential cross-section, we notice that leading-twist FFs, such as the Collins function $H^\perp_1$, are convoluted into the twist-3 effects. We therefore introduce two twist-3 azimuthal asymmetries, $\langle \cos\phi \rangle$ and $\langle \sin\varphi \rangle$, and provide corresponding numerical estimates to assess the feasibility of probing leading-twist FFs.  Numerical estimates indicate that these twist-3 effects are not significantly suppressed due to the contributions from weak interaction, and they could be measured at future high-energy electron–positron colliders \cite{CEPCStudyGroup:2018ghi,FCC:2018byv,FCC:2018evy}. This probably provides new opportunities for investigating twist-3 effects or FFs and exploring hadronization dynamics.

The paper is organized as follows. In Sec. \ref{sec:frame}, we define the center-of-mass system (CMS) of the leptons and introduce the kinematics for the fully inclusive jet production in the electron–positron annihilation. In Sec. \ref{sec:hadron}, we present a systematic framework for calculating the hadronic tensor and the differential cross-section up to twist-3 accuracy, incorporating both electromagnetic and weak interactions. To ensure gauge invariance, the quark–gluon–quark correlation function is explicitly included. In Sec. \ref{sec:asymmetry}, we introduce two twist-3 azimuthal asymmetries, $\langle \cos\phi \rangle$ and $\langle \sin\varphi \rangle$, and provide corresponding numerical estimates. Finally, a brief summary is presented in Sec. \ref{sec:summary}.

%Current theoretical research on jet production generally follows two paths: one focuses on higher-twist calculations \cite{Song:2010pf,Song:2013sja,Wei:2016far,Yang:2017sxz,Chen:2020ugq,Yang:2020qsk,Yang:2022knp,Yang:2022xwy,Yang:2023vyv,Yang:2023zod,Yang:2025hky}, while the other on higher-order perturbative corrections \cite{Gutierrez-Reyes:2018qez,Gutierrez-Reyes:2019vbx,Liu:2018trl,Liu:2020dct,Kang:2020fka,Arratia:2020ssx,Arratia:2019vju,Arratia:2022oxd}. 
%In the context of electron-positron annihilation, the study of FFs through fully inclusive jet production provides a particularly advantageous framework. By focusing on the final state of the jet, this approach simplifies the analysis compared to semi-inclusive processes that involve multiple hadrons. The clean kinematics of the annihilation process, coupled with the ability to measure a single jet, allow for a more precise and less parameter-dependent extraction of FFs.  To this end, we present 

\section{Reference Frame and Kinematic Setup}\label{sec:frame}

\begin{figure}
    \centering
    \includegraphics[width=0.9\linewidth]{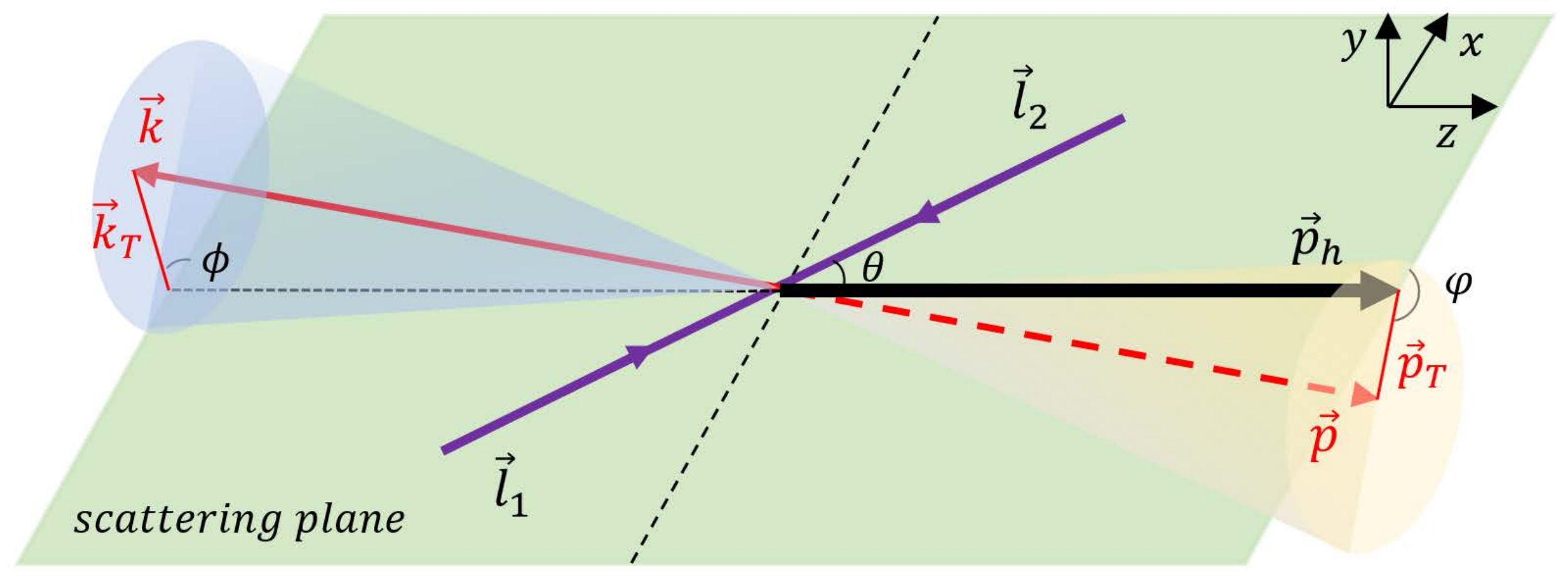}
    \caption{Illustration of the  center-of-mass system for the fully inclusive jet production electron-positron annihilation process. The angle  between the plane determined by $\vec{k}$ and $\vec{p}_h$ ($\vec{p}$ and $\vec{p}_h$) and the scattering plane is $\phi$ ($\varphi$). }
    \label{fig:cmsframe}
\end{figure}

Let us start by considering the reference frame and kinematics of fully inclusive jet production in electron-positron annihilation. We focus exclusively on the unpolarized case, assuming that both the initial leptons and the final state particles are unpolarized. We label this process as
\begin{align}
    e^-(l_1)+e^+(l_2)\to h(p_h) +jet (k) +X,
\end{align}
where momenta are shown in the parentheses.  We choose the center-of-mass system of the incoming leptons or the rest frame of the virtual gauge boson, see Fig. \ref{fig:cmsframe}. In this frame, the produced hadron travels along with the $+z$ direction and the three-momenta of the hadron ($\vec {p}_h$) and the incoming electron ($\vec{l}_1$) determine the $x-z$ plane or the scattering plane. The angle between $\vec {p}_h$ and $\vec{l}_1$ is $\theta$. The momentum of the jet is determined by $\vec{k}=\sum_i \vec{k}_i$, where $\vec{k}_i$ is the momentum of the $i$-th particle.  At leading order, quark and anti-quark are produced back-to-back in the center-of-mass system, i.e., $\vec{k}=-\vec{p}$, $\vec{p}$ is the momentum of the jet containing the measured hadron. The azimuthal angle between the plane determined by $\vec{k}$ and $\vec{p}_h$ ($\vec{p}$ and $\vec{p}_h$) and the scattering plane is $\phi$ ($\varphi$).

In the center-of-mass system, we can introduce the light-cone unit vectors, $\bar{n}^\mu=(1,0,\vec{0}_T)$ and $n^\mu=(0,1,\vec{0}_T)$, which satisfy $\bar{n}^2=n^2=0, \bar{n}\cdot n=1$. 
These vectors allow us to construct the transverse tensors for further analysis,
\begin{align}
    & g^{\mu\nu}_T=g^{\mu\nu}-\bar{n}^\mu n^\nu - \bar{n}^\nu n^\mu, \\
    & \varepsilon_T^{\mu\nu}=\varepsilon^{\mu\nu\alpha\beta}\bar{n}_\alpha n_\beta.
\end{align} 
We also introduce the following variables, 
\begin{align}
    &q^2=(l_1+l_2)^2, && y=\frac{p_h\cdot l_1}{p_h\cdot q}, && z=\frac{2p_h\cdot q}{Q^2}, 
\end{align}
and $s=Q^2=q^2$. In this case, we can parametrize the relevant momenta in terms of  unit light-cone vectors ($\bar n, n$).
\begin{align}
   & p_h^\mu = \frac{zQ}{\sqrt{2}}\bar n^\mu +\frac{M^2}{\sqrt{2}zQ}n^\mu,  \label{f:ph}\\
   & q^\mu= \frac{Q}{\sqrt{2}}\bar n^\mu +\frac{Q}{\sqrt{2}}n^\mu,  \label{f:q}\\
 %  & q^\mu= \frac{Q}{\sqrt{2}}\bar n^\mu +\frac{Q}{\sqrt{2}}n^\mu +q_T^\mu, \\
   & l_1^\mu= \frac{Q}{\sqrt{2}}(1-y)\bar n^\mu +\frac{Q}{\sqrt{2}}y n^\mu+l_T^\mu, \\
   & l_2^\mu= \frac{Q}{\sqrt{2}}y\bar n^\mu +\frac{Q}{\sqrt{2}}(1-y) n^\mu-l_T^\mu,   
\end{align}
Under the condition $M\ll Q$,  $p_h^\mu \sim zQ\bar n^\mu/\sqrt{2}$. The transverse momentum $l_T^\mu=Q\sqrt{y(1-y)}x^\mu$.  Momenta can also be parametrized in terms of the time-like vector $t^\mu$ and the space-like vector $v^\mu$. Relations between ($\bar n, n$) and ($t, v$) are presented in Appendix \ref{app:vector}.
% which satisfy $\bar n^2=n^2=0, \bar n\cdot n=1$. Since the virtual gauge boson is at rest, actually $q_T=0$ in this frame. It is also convenient to introduce the following tensors, 
%\begin{align}
 %   & g_T^\mu=g^{\mu\nu}-\bar n^\mu n^\nu- \bar n^\nu n^\mu, \\
%    & \varepsilon_T^{\mu\nu}= \varepsilon^{\alpha\beta\mu\nu} \bar n_\alpha n_\beta.
%\end{align}

We write down the differential cross section as 
\begin{align}
    \frac{d\sigma}{dzdyd\eta_Jd^2\vec{k}_T} =\frac{\alpha^2N_cz}{16\pi^2 sQ^4} A_r L^r_{\mu\nu}(l_1,l_2)W_r^{\mu\nu}(q,p_h),  \label{f:crosskine}
\end{align}
where $\eta_J$ is the rapidity of the measured jet, $\alpha$ is the fine structure constant, $N_c$ is the color factor, and $r=\gamma\gamma, \gamma Z, ZZ$ for electromagnetic, interference and weak terms, respectively. 
\begin{align}
    & A_{\gamma\gamma}= e_l^2 e_q^2, \\
    & A_{\gamma Z}= \frac{2e_l e_q Q^2(Q^2-M_Z^2)}{\left[(Q^2-M_Z^2)^2+\Gamma_Z M_Z^2\right]\sin^22\theta_W}, \\
    & A_{ZZ}= \frac{Q^4}{\left[(Q^2-M_Z^2)^2+\Gamma_Z M_Z^2\right]\sin^42\theta_W},
\end{align}
where $e_l, e_q$ are electric charges of the electron and the quark with flavour $q$, $\Gamma_Z$ and $M_Z$ are the width and the mass of the $Z^0$ boson, $\theta_W$ is the weak mixing angle. The leptonic tensors are given by
\begin{align}
    & L^{\gamma\gamma}_{\mu\nu}=4(l_{1\mu}l_{2\nu}+l_{1\nu}l_{2\mu}-g_{\mu\nu} l_1\cdot l_2), \\
    & L^{\gamma Z}_{\mu\nu}=4c_V^e(l_{1\mu}l_{2\nu}+l_{1\nu}l_{2\mu}-g_{\mu\nu} l_1\cdot l_2) -4ic_A^e \varepsilon_{\mu\nu l_1l_2}, \\
    & L^{ZZ}_{\mu\nu}=4c_1^e(l_{1\mu}l_{2\nu}+l_{1\nu}l_{2\mu}-g_{\mu\nu} l_1\cdot l_2) -4ic_3^e \varepsilon_{\mu\nu l_1l_2},
\end{align}
where $c_1^e=(c_V^e)^2+(c_A^e)^2$ and $c_3^e=2c_V^ec_A^e$. We note that $c_V^e, c_A^e$ are weak couplings and defined in the weak current $J_Z^\mu(0)=\bar\psi(0)\Gamma^\mu\psi(0)$ with $\Gamma^{\mu,e}=\gamma^\mu(c_V^e-c_A^e\gamma^5)$. The corresponding hadronic tensors are respectively written as 
\begin{align}
     W_{\gamma\gamma}^{\mu\nu}&= \sum_X\sum_i (2\pi)^3\delta^4(q-p_h-k_i-p_x)\nonumber \\
    &\times\langle 0|J_\gamma^\mu(0)|hX,i\rangle \langle hX,i|J_\gamma^\nu(0)|0\rangle, \\
     W_{\gamma Z}^{\mu\nu}&= \sum_X\sum_i (2\pi)^3\delta^4(q-p_h-k_i-p_x)\nonumber \\
    &\times\langle 0|J_\gamma^\mu(0)|hX,i\rangle \langle hX,i|J_Z^\nu(0)|0\rangle, \\
     W_{ZZ}^{\mu\nu}&= \sum_X\sum_i (2\pi)^3\delta^4(q-p_h-k_i-p_x)\nonumber \\
    &\times\langle 0|J_Z^\mu(0)|hX,i\rangle \langle hX,i|J_Z^\nu(0)|0\rangle. 
\end{align}
%Hadronic tensors are non-perturbative quantities and can be decomposed in terms of basic Lorentz tensors, e.g, Ref. \cite{Yang:2025hky}. However, in this paper, we focus on the systematical calculation of the differential cross-section in the framework of parton model, we therefore do not consider the kinematic decompositions of the hadronic tensors.

\section{Hadronic tensor and Cross section}\label{sec:hadron}

\begin{figure}
    \centering
    \includegraphics[width=0.98\linewidth]{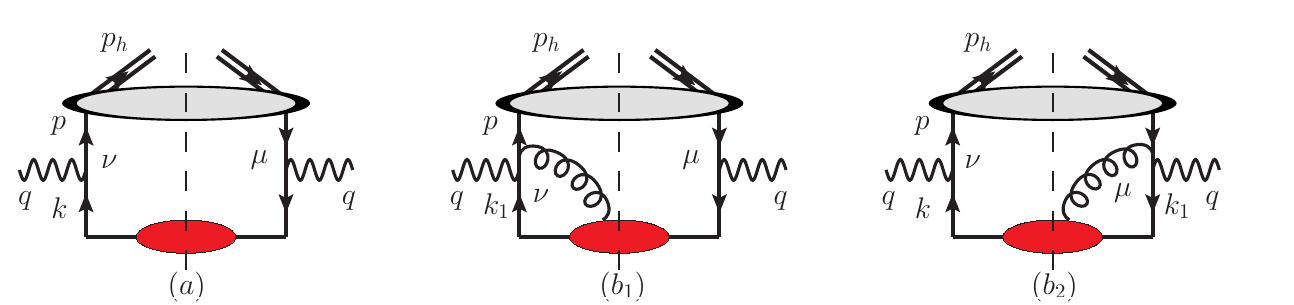}
    \caption{Feynman diagrams for the hadronic tensor. ($a$) denotes the handbag diagram without gluon scattering while ($b_1$) and ($b_2$) denote the one-gluon scattering diagrams.}
    \label{fig:jetxi}
\end{figure}

In the parton model, the hadronic tensor up to twist-3 accuracy can be calculated using Feynman diagrams (see Fig. \ref{fig:jetxi}). Diagram ($a$) represents the handbag contribution, which involves only quark-quark correlators without gluon exchange.
It serves as the basic leading-twist contribution to the hadronic tensor. Diagrams ($b_1$) and ($b_2$), on the other hand, include quark-gluon-quark jet correlators (bottom block), showing the contributions from one-gluon scattering processes. 
It should be clarified that the present work focuses on the twist-3 effects arising from the fully inclusive jet correlators, while the kinematic and dynamical higher-twist contributions from the fragmentation correlator are neglected. Therefore, our results represent a specific, rather than a fully comprehensive, twist-3 calculation. A complete treatment of twist-3 effects, including the structures of the fragmentation side, has been extensively investigated in Refs. \cite{Boer:1997mf,Bacchetta:2006tn,Chen:2016moq}. In this case, the hadronic tensor can be written as
%It is important to note that the complete twist-3 hadronic tensor should also include the quark-gluon-quark fragmentation correlator (one gluon attached to the top block) which generates the twist-3 FFs \cite{Boer:1997mf,Bacchetta:2006tn,Chen:2016moq}. However, since our aim is to study twist-3 effects arising from fully inclusive jets, we focus on the leading-twist FFs by considering only the contributions from the quark–gluon–quark jet correlators. 
\begin{align}
    W^{\mu\nu}=\int &\frac{d^2\vec p_Td^2\vec k_T}{(2\pi)^2}\delta^2(\vec k_T+\vec p_T)\bigg\{ 
    \mathrm{Tr}\left[\hat{J}(k_T)\Gamma^\mu \hat\Xi(z,p_T)\Gamma^\nu\right] \nonumber \\
    &-\frac{1}{\sqrt{2}Q}\mathrm{Tr} \left[\hat{J}^{\alpha,L}(k_T)\Gamma^\mu \slashed n \gamma^\alpha\hat\Xi(z,p_T)\Gamma^\nu\right] \nonumber \\
    &-\frac{1}{\sqrt{2}Q}\mathrm{Tr} \left[\hat{J}^{\alpha,R}(k_T)\Gamma^\mu\hat\Xi(z,p_T)  \gamma^\alpha\slashed n\Gamma^\nu\right]\bigg\},
\end{align}
where $\hat{J}(k_T), \hat{J}^{\alpha}(k_T)$ are jet correlators which are respectively defined as 
\begin{align}
     \hat{J}(k_T)=&\int d\eta^+d^2\eta_Te^{ik^-\eta^+-i\vec k_T\vec\eta_T}\nonumber \\
    &\times\langle 0|\bar\psi(0)\mathcal{L}(0,\eta)\psi(\eta)|0\rangle, \\
     \hat{J}^{\alpha,L}(k_T)=&\int d\eta^+d^2\eta_Te^{ik^-\eta^+-i\vec k_T\vec\eta_T}\nonumber \\
    &\times\langle 0|\bar\psi(0)\mathcal{L}(0,\eta)D_T^\alpha(\eta)\psi(\eta)|0\rangle.
\end{align}
Superscripts $R, L$ denote the right-cut and the left-cut, respectively, which also correspond to $(b_1)$ and $(b_2)$ in Fig. \ref{fig:jetxi}.
The fragmentation correlator $\hat\Xi(z,p_T)$ is given by
\begin{align}
    \hat\Xi(z,p_T)=&\frac{1}{2\pi}\sum_X\int d\xi^+d^2\xi_Te^{ip_h^+\xi^-/z-i\vec p_T\vec\xi_T} \nonumber \\
    &\times\langle 0|\psi(0)\mathcal{L}(0,\infty)|hX\rangle \langle hX|\mathcal{L}(\infty,\xi)\bar\psi(\xi)|0\rangle.
\end{align}
Gauge link $\mathcal{L}(0,y)$ has been inserted to keep the correlator gauge invariance. 

To calculate the hadronic tensor, we decompose these correlators in terms of $\Gamma$-matrices.  The fragmentation correlator can be decomposed as
\begin{align}
     z\hat{\Xi}(z, p_T)=\slashed{\bar{n}} D_1(z,p_T) +i\sigma^{\mu\nu}\gamma^5 \bar{n}_\mu \frac{\varepsilon_{T~\nu}^p }{M}H^\perp_1(z,p_T),
\end{align}
at leading-twist accuracy for unpolarized hadron production. For the quark-quark jet correlator, we decompose it in the following form,
\begin{align} 
     \hat{J}(k^-,k_T)&=\frac{\slashed n}{2} \alpha_1(k^-,k_T) +i\sigma^{\alpha\beta}\gamma^5  \frac{n_\alpha\varepsilon^k_{T~\beta}}{2M_J}\omega^\perp_1(k^-,k_T) \nonumber   \\
    &+ \gamma^5\gamma_\alpha\frac{ \varepsilon_T^{k\alpha} }{2k^-}\beta^\perp(k^-,k_T) + \frac{\slashed k_T}{2k^-} \alpha^\perp(k^-,k_T) \nonumber   \\
    &+ \frac{M_J}{2k^-}\epsilon(k^-,k_T) + i\sigma^{\alpha\beta}\gamma^5\varepsilon_{T\alpha\beta} \frac{M_J}{2k^-}\omega(k^-,k_T),
    \label{f:jetfunction}
\end{align}
where $\alpha, \beta, \epsilon$ and $\omega$ are scalar coefficient functions. % For simplicity, arguments ($k^-, k_T$) are omitted. 
We notice that our decomposition is a little different from that in Ref. \cite{Accardi:2020iqn}, Eq. \eqref{f:jetfunction} includes parity-violating terms to account for the contributions from the weak interaction. As a first step, we decompose the quark-gluon-quark correlator into the following form,
\begin{align}
     \hat{J}^{\alpha,L}(k^-,k_T)&=-\frac{\slashed n}{2} k_T^\alpha \alpha_d^\perp(k^-,k_T)+i\frac{\gamma^5\slashed n}{2} \varepsilon_T^{k\alpha}\beta_d^\perp(k^-,k_T)  \nonumber \\
      &+ \frac{i}{4}\gamma^\rho\varepsilon^\alpha_{T~\rho}\gamma^5 \slashed{n} M_J\epsilon(k^-,k_T) \nonumber \\
      &+\frac{ i}{2}\sigma^{\rho\sigma}\gamma^5 n_\sigma\varepsilon^\alpha_{T~\rho}M_J\omega(k^-,k_T). 
%     \hat{J}^{\alpha,L}(k_T)&=-\frac{\slashed n}{2} k_T^\alpha \alpha_d^\perp+i\frac{\gamma^5\slashed n}{2} \varepsilon_T^{k\alpha}\beta_d^\perp-\frac{1}{4}\sigma^{\rho\sigma}\gamma^5 n_\sigma\varepsilon^\alpha_{T~\rho}M_J\epsilon \nonumber \\
%      &+\frac{ i}{2}\sigma^{\rho\sigma}\gamma^5 n_\sigma\varepsilon^\alpha_{T~\rho}M_J\omega.
\end{align}
We would like to highlight here that the introduction of the quark-gluon-quark correlator is motivated by two key factors.  First, it serves as one of the dynamical origins of twist-3 contributions. Second, it ensures that the hadronic tensor satisfies current conservation at the twist-3 level.
%, i.e.,  $q_\mu W^{\mu\nu}=q_\nu W^{\mu\nu}=0$.
%where $2\omega_d=2\omega+i\epsilon$. 

\subsection{Chiral-even contributions}

In the following, we demonstrate how to systematically calculate the hadronic tensor. To make the procedure clearer, we divide the calculation into two parts: the chiral-even part and the chiral-odd part. We first consider the chiral-even contribution.

The leading-twist contribution only originates from the handbag diagram. Therefore, we can calculate the trace,
\begin{align}
  \mathrm{Tr}\left[\slashed n \Gamma^\mu \slashed{\bar{n}}\Gamma^\nu\right]=-\left[4c_1^qg_T^{\mu\nu} + 4ic_3^q\varepsilon_T^{\mu\nu}\right]
\end{align}
to obtain the corresponding hadronic tensor. It is given by 
\begin{align}
  %W^{\mu\nu}_2= {\int\hspace{-0.9em}2}_{pk} \Bigg\{-\left[4c_1^qg_T^{\mu\nu} + 4ic_3^q\varepsilon_T^{\mu\nu}\right]\frac{\alpha_1 D_1}{z}\Bigg\}, 
  W^{\mu\nu}_{t2,e}= \mathcal{C} \Bigg\{-\left[2c_1^qg_T^{\mu\nu} + 2ic_3^q\varepsilon_T^{\mu\nu}\right]\frac{\alpha_1 D_1}{z}\Bigg\},  \label{f:w2}
\end{align}
where we have used short-handed notation 
\begin{align}
    \mathcal{C}\{ ~ ~ \}=\int \frac{d^2p_Td^2k_T}{(2\pi)^2}\delta^2(\vec k_T+\vec p_T)\Bigg\{ ~ ~ \Bigg\}
\end{align}
 to simplify the expression. The subscripts $t2$ and $e$ denote the leading twist and chiral-even hadronic tensor.

There are two distinct sources for the twist-3 hadronic tensor, one is the contribution from the quark-quark correlator  $\hat{J}(k_T)$, and the other is from the quark-gluon-quark correlator  $\hat{J}^\alpha(k_T)$. Considering the former first, we use the following traces,
\begin{align}
    & \mathrm{Tr}\left[\gamma^\alpha \Gamma^\mu \slashed{\bar{n}}\Gamma^\nu\right]k_{T\alpha}=\left[4c_1^qk_T^{\{\mu}\bar{n}^{\nu\}} + 4ic_3^q\varepsilon_T^{k[\mu}\bar{n}^{\nu]}\right], \\
    & \mathrm{Tr}\left[\gamma^5\gamma^\alpha \Gamma^\mu \slashed{\bar{n}}\Gamma^\nu\right]\varepsilon_{T~\alpha}^k=-\left[4c_3^q\varepsilon_T^{k\{\mu}\bar{n}^{\nu\}} - 4ic_1^q k_T^{[\mu}\bar{n}^{\nu]}\right], 
\end{align}
 and obtain the contribution from $\hat{J}(k_T)$. We write the hadronic tensor as 
\begin{align}
  W^{\mu\nu,0}_{t3,e}= &\mathcal{C} \Bigg\{\left[2c_1^qk_T^{\{\mu}\bar{n}^{\nu\}} + 2ic_3^q\varepsilon_T^{k[\mu}\bar{n}^{\nu]}\right]\frac{\alpha^\perp D_1}{zk^-} \nonumber \\
  & -\left[2c_3^q\varepsilon_T^{k\{\mu}\bar{n}^{\nu\}} - 2ic_1^q k_T^{[\mu}\bar{n}^{\nu]}\right] \frac{\beta^\perp D_1}{zk^-}\Bigg\}.\label{f:w3j2}
\end{align}
We have  used notations $A^{\{\mu}B^{\nu\}}=A^\mu B^\nu +A^\nu B^\mu$, $A^{[\mu}B^{\nu]}=A^\mu B^\nu -A^\nu B^\mu$ to simplify the results.

To obtain the twist-3 contribution from $\hat{J}^\alpha(k_T)$, which corresponds to $(b_1)$ in Fig. \ref{fig:jetxi}, we use the following traces,
\begin{align}
    & \mathrm{Tr}\left[\slashed n \Gamma^\mu \slashed n\gamma^\alpha\slashed{\bar{n}}\Gamma^\nu\right]k_{T\alpha}=-\left[8c_1^qk_T^{\nu}n^{\mu} - 8ic_3^q\varepsilon_T^{k\nu}n^{\mu}\right], \\
    & \mathrm{Tr}\left[\gamma^5\slashed n \Gamma^\mu \slashed n\gamma^\alpha\slashed{\bar{n}}\Gamma^\nu\right]i\varepsilon_{T~\alpha}^k=-\left[8c_1^qk_T^{\nu}n^{\mu} - 8ic_3^q\varepsilon_T^{k\nu}n^{\mu}\right], 
\end{align}
and obtain the hadronic tensor as
\begin{align}
  W^{\mu\nu,L}_{t3,e}= &\mathcal{C} \Bigg\{-\left[4c_1^qk_T^{\nu}n^{\mu} - 4ic_3^q\varepsilon_T^{k\nu}n^{\mu}\right]\frac{\alpha^\perp D_1}{\sqrt{2}zQ} \nonumber \\
  & +\left[4c_3^q\varepsilon_T^{k\nu}n^{\nu} +4ic_1^q k_T^{\nu}n^{\mu}\right] \frac{\beta^\perp D_1}{\sqrt{2}zQ}\Bigg\}. \label{f:w3j3L} 
\end{align}
Here, we have used the relation, $\alpha_d^\perp-\beta^\perp_d=\alpha^\perp-i\beta^\perp$, which follows from the equation of motion for massless QCD, $\slashed D\psi(x)=0$ \cite{Chen:2016moq,Yang:2022knp}. The complete result from the quark–gluon–quark correlator includes contributions from both $(b_1)$ and $(b_2)$ in Fig. \ref{fig:jetxi}, i.e., $ W^{\mu\nu,L}_{t3,e}+( W^{\nu\mu,L}_{t3,e})^*$. Therefore, we have 
\begin{align}
   &W^{\mu\nu,L}_{t3,e}+( W^{\nu\mu,L}_{t3,e})^* \nonumber \\
   = &\mathcal{C} \Bigg\{-\left[2c_1^qk_T^{\{\mu}q^-n^{\nu\}} +2ic_3^q\varepsilon_T^{k[\mu}q^-n^{\nu]}\right]\frac{\alpha^\perp D_1}{p_h\cdot q} \nonumber \\
  & \quad +\left[2c_3^q\varepsilon_T^{k\{\mu}q^- n^{\nu\}} -2ic_1^q k_T^{[\mu}q^- n^{\nu]}\right] \frac{\beta^\perp D_1}{p_h\cdot q}\Bigg\}.\label{f:w3j3}
\end{align}

The complete twist-3 hadronic tensor for the chiral-even part is then given by the sum of $  W^{\mu\nu,0}_{t3,e}$ and $W^{\mu\nu,L}_{t3,e}+( W^{\nu\mu,L}_{t3,e})^*$.  Finally, we obtain
\begin{align}
  W^{\mu\nu}_{t3,e}&= W^{\mu\nu,0}_{t3} +W^{\mu\nu,L}_{t3}+( W^{\nu\mu,L}_{t3})^*   \nonumber \\
  &=\mathcal{C} \Bigg\{-\left[2c_1^qk_T^{\{\mu}\bar{q}^{\nu\}} + 2ic_3^q\varepsilon_T^{k[\mu}\bar{q}^{\nu]}\right]\frac{\alpha^\perp D_1}{p_h\cdot q} \nonumber \\
  &\quad \ \quad +\left[2c_3^q\varepsilon_T^{k\{\mu}\bar{q}^{\nu\}} - 2ic_1^q k_T^{[\mu}\bar{q}^{\nu]}\right] \frac{\beta^\perp D_1}{p_h\cdot q}\Bigg\},  \label{f:w3}
\end{align}
where $\bar{q}^\mu=q^\mu-2p_h^\mu/z$ and it satisfies $q\cdot \bar{q}=0$. The leads to the hadronic tensor $W^{\mu\nu}_{t3}$ to be gauge invariant, $q_\mu W_{t3,e}^{\mu\nu}=q_\nu W_{t3,e}^{\mu\nu}=0$.  We also note that $\bar{q}^\mu = - Qv^\mu$, $v^\mu$ is defined as the space-like vector and introduced in Appendix \ref{app:vector}.

\subsection{Chiral-odd contributions}

Now let us turn our attention to the chiral-odd part. Following the same method, we first calculate the leading-twist contribution, followed by the twist-3 contribution. Using the following trace,
\begin{align}
  &\mathrm{Tr}\left[i\sigma^{\rho\sigma}\gamma^5\Gamma^\mu i\sigma^{\alpha\beta}\gamma^5\Gamma^\nu \right]n_\rho \varepsilon_{T~\sigma}^k \bar{n}_\alpha\varepsilon_{T~\beta}^p \nonumber \\
  &=4c_2^q\left[p_{T}^{\{\mu}k_T^{\nu\}}-g_T^{\mu\nu} p_T\cdot k_T\right],
\end{align}
we obtain the leading-twist chiral-odd hadronic tensor,
\begin{align}
    W^{\mu\nu}_{t2,o}&= \mathcal{C} \Bigg\{2c_2^q\left[p_{T}^{\{\mu}k_T^{\nu\}}-g_T^{\mu\nu} p_T\cdot k_T\right]\frac{\omega^\perp_1 H^\perp_1}{zM_JM} \Bigg\}. \label{f:t2}
\end{align}
The subscript  $o$ denotes the chiral-odd hadronic tensor. The coefficient $c_2^q=(c_V^q)^2-(C_A^q)^2$. 

For the twist-3 contribution from the quark-quark correlator $\hat{J}(k_T)$, we use
\begin{align}
  &\mathrm{Tr}\left[i\sigma^{\rho\sigma}\gamma^5\Gamma^\mu i\sigma^{\alpha\beta}\gamma^5\Gamma^\nu \right]\varepsilon_{T\rho\sigma}^k \bar{n}_\alpha\varepsilon_{T~\beta}^p =-8c_2^q p_{T}^{\{\mu}\bar{n}^{\nu\}}, \nonumber  \\
  & \mathrm{Tr}\left[\Gamma^\mu i\sigma^{\alpha\beta}\gamma^5\Gamma^\nu \right] \bar{n}_\alpha\varepsilon_{T~\beta}^p =4ic_2^q p_{T}^{[\mu}\bar{n}^{\nu]},
\end{align}
and obtain 
\begin{align}
  W^{\mu\nu,0}_{t3,o}&=  \mathcal{C} \Bigg\{2ic_2^q p_{T}^{[\mu}\bar{n}^{\nu]} \frac{M_J}{M}\frac{\epsilon H^\perp_1}{zk^-} -4c_2^q p_{T}^{\{\mu}\bar{n}^{\nu\}} \frac{M_J}{M}\frac{\omega H^\perp_1}{zk^-}  \Bigg\}.  \label{f:t3j2}
\end{align}

For the twist-3 contribution from the quark-gluon-quark correlator $\hat{J}^\alpha(k_T)$, we use
\begin{align}
  &\mathrm{Tr}\left[\sigma^{\rho\sigma}\gamma^5\Gamma^\mu \slashed n \gamma^\tau  \sigma^{\alpha\beta}\gamma^5\Gamma^\nu \right]n_\sigma\varepsilon_{T\tau\rho}  \bar{n}_\alpha \varepsilon_{T~\beta}^p =16c_2^q p_{T}^{\nu}n^{\mu}, \label{f:t3j3}
%  &\mathrm{Tr}\left[i\sigma^{\rho\sigma}\gamma^5\Gamma^\mu \slashed n \gamma^\tau  i\sigma^{\alpha\beta}\gamma^5\Gamma^\nu \right]n_\sigma\varepsilon_{T\tau\rho}  \bar{n}_\alpha \varepsilon_{T~\beta}^p =-16c_2^q p_{T}^{\nu}n^{\mu}, \label{f:t3j3}
\end{align}
and obtain
\begin{align}
  W^{\mu\nu,L}_{t3,o}=  \mathcal{C} \Bigg\{2ic_2^q p_{T}^{\nu}n^{\mu} \frac{M_J}{M}\frac{\epsilon H^\perp_1}{p_h\cdot q}  +4c_2^q p_{T}^{\nu} n^{\mu} \frac{M_J}{M}\frac{\omega H^\perp_1}{p_h\cdot q} \Bigg\}. 
\end{align}
%Therefore, $ W^{\mu\nu,L}_{t3,o}+( W^{\nu\mu,L}_{t3,o})^*$ is given by 
Considering the result from the quark-gluon-quark correlator  includes contributions from both $(b_1)$ and $(b_2)$ in Fig. \ref{fig:jetxi}, we obtain
\begin{align}
   &W^{\mu\nu,L}_{t3,o}+( W^{\nu\mu,L}_{t3,o})^* \nonumber \\
   = & \mathcal{C} \Bigg\{2ic_2^q p_{T}^{[\nu}n^{\mu]} \frac{M_J}{M}\frac{\epsilon H^\perp_1}{p_h\cdot q}  +4c_2^q p_{T}^{\{\nu} n^{\mu\}} \frac{M_J}{M}\frac{\omega H^\perp_1}{p_h\cdot q} \Bigg\}.  \label{f:wt3o}
\end{align}

Similarly to Eq. \eqref{f:w3j3}, the complete twist-3 hadronic tensor for chiral-odd part is  the sum of $ W^{\mu\nu,0}_{t3,o}$ and $W^{\mu\nu,L}_{t3,o}+( W^{\nu\mu,L}_{t3,o})^*$. We therefore obtain
\begin{align}
  W^{\mu\nu}_{t3,o}&=   W^{\mu\nu,0}_{t3,o}+ W^{\mu\nu,L}_{t3,o}+( W^{\nu\mu,L}_{t3,o})^* \nonumber \\
   =&\mathcal{C} \Bigg\{-2ic_2^q p_{T}^{[\nu}\bar{q}^{\mu]} \frac{M_J}{M}\frac{\epsilon H^\perp_1}{p_h\cdot q}  +4c_2^q p_{T}^{\{\nu} \bar{q}^{\mu\}} \frac{M_J}{M}\frac{\omega H^\perp_1}{p_h\cdot q} \Bigg\}.  \label{f:wt3}
\end{align}
We can see that Eq. \eqref{f:wt3} also  satisfies the current conservation,  $q_\mu W_{t3,o}^{\mu\nu}=q_\nu W_{t3,o}^{\mu\nu}=0$.

At the end of this subsection, we present the complete hadronic tensor up to twist-3. It is given by
\begin{align}
    W^{\mu\nu}=\mathcal{C} \Bigg\{&-\left[2c_1^qg_T^{\mu\nu} + 2ic_3^q\varepsilon_T^{\mu\nu}\right]\frac{\alpha_1 D_1}{z} \nonumber  \\
    & -\left[2c_1^qk_T^{\{\mu}\bar{q}^{\nu\}} + 2ic_3^q\varepsilon_T^{k[\mu}\bar{q}^{\nu]}\right]\frac{\alpha^\perp D_1}{p_h\cdot q} \nonumber \\
   & +\left[2c_3^q\varepsilon_T^{k\{\mu}\bar{q}^{\nu\}} - 2ic_1^q k_T^{[\mu}\bar{q}^{\nu]}\right] \frac{\beta^\perp D_1}{p_h\cdot q} \nonumber \\
   &+2c_2^q\left[p_{T}^{\{\mu}k_T^{\nu\}}-g_T^{\mu\nu} p_T\cdot k_T\right]\frac{\omega^\perp_1 H^\perp_1}{zM_JM} \nonumber \\
   &-2c_2^q \left[ip_{T}^{[\nu}\bar{q}^{\mu]} \frac{\epsilon H^\perp_1}{p_h\cdot q}  +2 p_{T}^{\{\nu} \bar{q}^{\mu\}} \frac{\omega H^\perp_1}{p_h\cdot q}\right]\frac{M_J}{M}  \Bigg\}. 
\end{align}
Here, we present the hadronic tensor for the weak interaction. To obtain the hadronic tensor for the electromagnetic interaction or the interference terms, appropriate substitutions are required, see Table \ref{tab:replacing}. We note that the complete hadronic tensor includes all three contributions.

\begin{table}
\renewcommand\arraystretch{1.5}
\begin{tabular}{cccc}
\hline \hline
~~~  & $A_r$  & $L_{\mu\nu}^r$  & $W^{\mu\nu}_r$  \\ \hline 
~ $ZZ$~ & $A_{ZZ}$ & $c_1^e,~c_3^e$ & $c_1^q,~c_3^q$  \\
$\gamma Z$  & ~$A_{ZZ}\to A_{\gamma Z}$ ~& ~~$c_1^e\to c_V^e,~c_3^e\to c_A^e$~ &~~ $c_1^q\to c_V^q,~c_3^q\to c_A^q$~~ \\
$\gamma\gamma$ & $A_{ZZ}\to A_{\gamma \gamma}$  & $c_1^e\to 1,~c_3^e\to 0$  & $c_1^q\to 1,~c_3^q\to 0$ \\ \hline \hline
\end{tabular}
\caption{Relations of kinematic factors between weak, EM and interference interactions.}
\label{tab:replacing}
\end{table}

\subsection{Differential cross section}

The differential cross-section is obtained by contracting the leptonic tensor with the hadronic tensor, as given in Eq. \eqref{f:crosskine}. Here, we list the relevant contractions
\begin{align}
  & L_{\mu\nu}\cdot \left[c_1^qg_T^{\mu\nu} + ic_3^q\varepsilon_T^{\mu\nu}\right] =-4Q^2T_0(y), \\
  & L_{\mu\nu}\cdot \left[c_1^qk_T^{\{\mu}\bar{q}^{\nu\}} + ic_3^q\varepsilon_T^{k[\mu}\bar{q}^{\nu]}\right] =8Q^3T_1(y)|k_T|\cos\phi, \\
  & L_{\mu\nu}\cdot \left[c_3^q\varepsilon_T^{k\{\mu}\bar{q}^{\nu\}} - ic_1^q k_T^{[\mu}\bar{q}^{\nu]}\right] =8Q^3T_2(y)|k_T|\sin\phi, \\
  & L_{\mu\nu}\cdot c_2^q\left[p_{T}^{\{\mu}k_T^{\nu\}}-g_T^{\mu\nu} p_T\cdot k_T\right]= -8Q^2t_0(y)|p_T||k_T|\cos(\varphi+\phi), \\
  & L_{\mu\nu}\cdot c_2^q p_{T}^{\{\nu} \bar{q}^{\mu\}} =8Q^3t_1(y)|p_T|\cos\varphi, \\
  & L_{\mu\nu}\cdot c_2^q i p_{T}^{[\nu}\bar{q}^{\mu]} =8Q^3t_2(y)|p_T|\sin\varphi.
\end{align}
Here, the functions T(y) and t(y) are defined as 
\begin{align}
  & T_0(y)=c_1^ec_1^q A(y)-c_3^ec_3^qB(y), \\
  & T_1(y)=c_1^ec_1^q C(y)+c_3^ec_3^qD(y), \\  
  & T_2(y)=c_1^ec_3^q C(y)+c_3^ec_1^qD(y), \\
  & t_0(y)=c_1^ec_2^q y(1-y), \\
  & t_1(y)=c_1^ec_2^q C(y), \\
  & t_2(y)=c_3^ec_2^q D(y),
\end{align}
with
\begin{align}
  & A(y)=2y^2-2y+1, \\
  & B(y)=1-2y, \\
  & C(y)=(1-2y)\sqrt{y(1-y)}, \\
  & D(y)=\sqrt{y(1-y)}. 
\end{align}
Finally, the differential cross-section can be written as 
\begin{align}
    & d\tilde{\sigma}=\frac{\alpha^2 N_c}{2\pi^2Q^4}A_r \mathcal{C} \Bigg\{  T_0(y)\alpha_1D_1 -T_1(y)|k_T|\cos\phi \frac{4\alpha^\perp D_1}{Q} \nonumber\\
    &+T_2(y)|k_T|\sin\phi \frac{4\beta^\perp D_1}{Q} - t_0(y)|p_T||k_T|\cos(\varphi+\phi)\frac{2\omega^\perp_1 H^\perp_1}{M_JM}\nonumber\\
    &+t_1(y)|p_T|\cos\varphi \frac{\omega H^\perp_1}{Q}\frac{M_J}{M} - t_2(y)|p_T|\sin\varphi \frac{\epsilon H^\perp_1}{Q}\frac{M_J}{M} \Bigg\}. \label{f:corssfinal}
\end{align}
where $d\tilde{\sigma}=d\sigma/zdyd\eta_Jd^2\vec{k}_T$. We notice that the $T_0(y)$ term and the $t_0(y)$ term are leading-twist effects, other terms are twist-3 effects. Also, $T_2(y)$ term and $t_2(y)$ vanish if only EM interaction is taken into consideration.

\section{Twist-3 effects} \label{sec:asymmetry}

According to the differential cross-section in Eq. \eqref{f:corssfinal}, it is possible to isolate four twist-3 terms and subsequently extract the corresponding FFs ($D_1$, $H_1^\perp$) by measuring different azimuthal modulations. This offers a helpful demonstration of how leading-twist FFs might be accessed through twist-3 effects. Once these coefficient functions ($\alpha, \beta, \epsilon, \omega$) are properly identified, the transverse momentum dependent (TMD) FFs can be obtained.
Notably, $\alpha$ and $\epsilon$ can be explicitly determined as $\alpha=\alpha^\perp=1$ and $\epsilon=1$, with the detailed derivation provided in Appendix \ref{app:func}. Based on this result, we now discuss the potential for investigating leading-twist FFs through twist-3 effects. In the following section, we examine two specific twist-3 azimuthal asymmetries, $\langle \cos\phi \rangle$ and $\langle \sin\varphi \rangle$, which are directly related to $\alpha^\perp$ and $\epsilon$, respectively.

\begin{figure}
  \centering
  \includegraphics[width=0.9\linewidth]{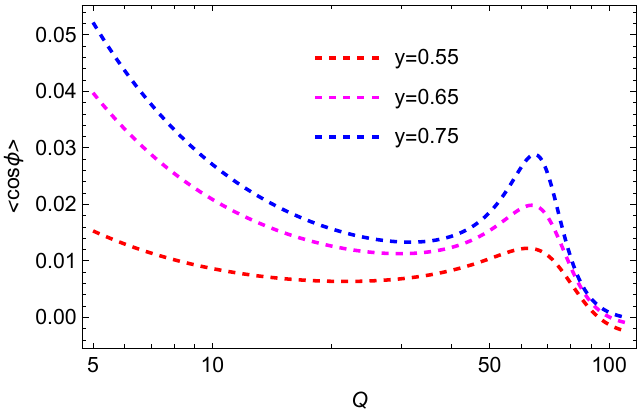}
  \caption{Numerical estimates of the azimuthal asymmetry $\langle \cos\phi \rangle$  with respect to Q.  The fraction $z$ is taken as $z=0.3$.}\label{fig:cosQ}
\end{figure}

\begin{figure}
  \centering
  \includegraphics[width=0.9\linewidth]{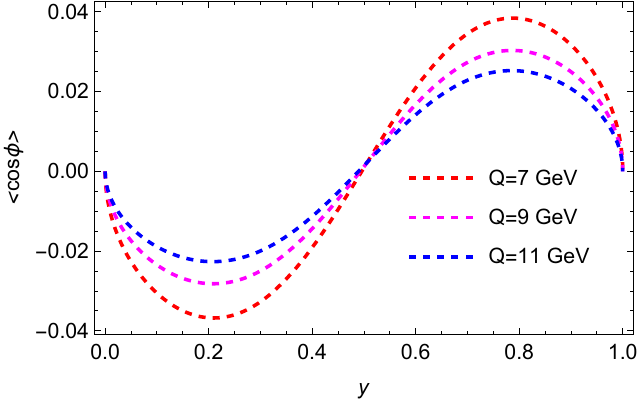}
  \caption{Numerical estimates of the azimuthal asymmetry $\langle \cos\phi \rangle$  with respect to y. The fraction $z$ is taken as $z=0.3$.}\label{fig:cosy}
\end{figure}

\begin{figure}
  \centering
  \includegraphics[width=0.9\linewidth]{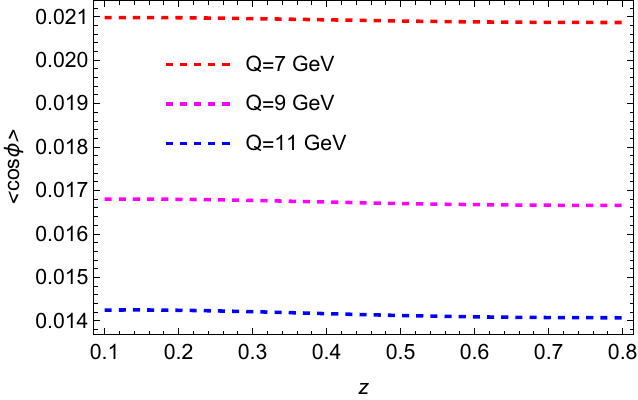}
  \caption{Numerical estimates of the azimuthal asymmetry $\langle \cos\phi \rangle$  with respect to z. The kinematic factor $y$ is taken as $y=0.6$.}\label{fig:cosz}
\end{figure}

We use the following formula to define azimuthal asymmetry $\langle \cos\phi \rangle$, 
\begin{align}
   \langle \cos\phi \rangle &=\frac{\int d\tilde{\sigma}\cos\phi d\phi d\varphi}{\int d\tilde{\sigma} d\phi d\varphi}.
\end{align} 
According to the differential cross-section in Eq. \eqref{f:corssfinal}, we have 
\begin{align}
  \langle \cos\phi \rangle &=-\frac{A_r4 \mathcal{C} \left\{ T_1(y)|k_T|\alpha^\perp D_1(z, p_T)\right\}}{A_r Q \mathcal{C} \left\{T_0(y)\alpha_1D_1(z, p_T)\right\}}.  \label{f:cosphi}
\end{align}
It should be noted that a summation over quark flavors is implied in both the numerator and the denominator, which also applies to similar equations below.
To deal with the TMD FF $D_1(z, p_T)$, we use the Gaussian ansatz, which is given by
\begin{align}
  D_1(z, p_T)=\frac{1}{\pi \Delta^2}e^{-\vec{p}_T^2/\Delta^2} D_1(z).
\end{align} 
We substitute function $D_1(z, p_T)$ into Eq. \eqref{f:cosphi} and obtain the simplified result, 
\begin{align}
   \langle \cos\phi \rangle &=- \frac{\Delta\sqrt{\pi}}{Q}\frac{A_rT_1(y) D_1(z)}{A_r T_0(y)D_1(z)}.  \label{f:cos}
\end{align}
In obtaining this result, the conditions $\alpha^\perp=1$ and $\epsilon=1$ have been applied. We note that $\langle \cos\phi \rangle$ receives contributions from both electromagnetic and weak interactions. To evaluate its magnitude and provide a clearer physical intuition of this asymmetry, and further assess the feasibility of extracting the FFs, we present numerical estimates in Figs. \ref{fig:cosQ}, \ref{fig:cosy}, and \ref{fig:cosz} as functions of $Q$, $y$, and $z$, respectively.
Fragmentation functions $D_1(z)$ are taken from NPC23 \cite{Gao:2024nkz,Gao:2024dbv} for $\pi^+$ meson. All light flavors ($u, \bar{u},  d, \bar{d}, s, \bar{s}$) are taken into account, including both favored and disfavored quarks.  The mean squared transverse momenta of TMD FFs are taken as $\Delta^2=0.17$ GeV$^2$ for all flavors \cite{Anselmino:2005nn,Signori:2013mda,Anselmino:2013lza,Cammarota:2020qcw,Bacchetta:2022awv,Bacchetta:2024qre}.

\begin{figure}
  \centering
  \includegraphics[width=0.9\linewidth]{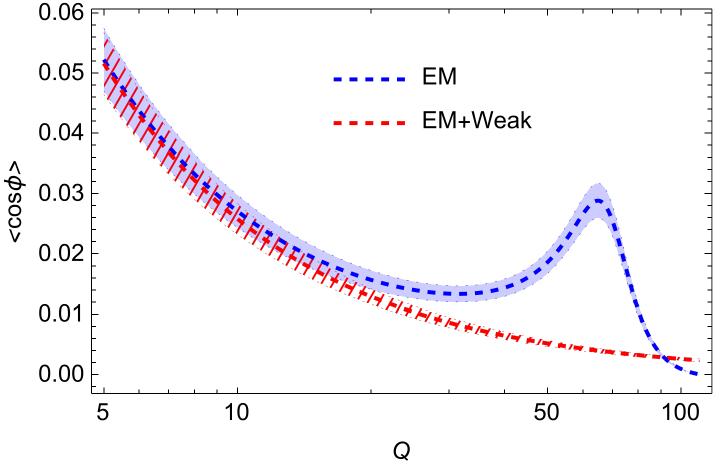}
  \caption{Numerical estimates of azimuthal asymmetry $\langle \cos\phi \rangle$  with respect to Q. We have used $z=0.3$ and $y=0.75$. The red dashed (blue dot-dashed) line shows the estimate for contributions of the electromagnetic interaction (both the electromagnetic and weak interactions). The bands stem from the sensitivity analysis by varying parameters by $\pm10\%$ uncertainties.}\label{fig:cosEW}
\end{figure}

Under the condition of Gaussian ansatz for the TMD FFs, numerical estimates lead to  the following observations.
\begin{itemize}
  \item As illustrated in Fig. \ref{fig:cosQ}, the azimuthal asymmetry is increasingly suppressed at higher energy scales due to its twist-3 nature.  Considering that the asymmetry contains the parity-violating component which is enhanced by weak interaction,  a competition thus arises between these two mechanisms. Nevertheless, at high energy scales, the contribution from weak interaction becomes dominant. In Fig. \ref{fig:cosEW}, we present a comparison between the asymmetry contributions arising from pure EM interaction and those including both EM and weak interaction effects. The bands predominantly reflect the sensitivity analysis where parameters are varied by $\pm 10\%$ uncertainties, whereas the uncertainties originating from the FFs are negligible. We can see that the weak interaction contribution tends to be noticeable at energies above 10 GeV. This implication points to the potential feasibility of detecting $\langle \cos\phi \rangle$ at future high-energy electron-positron facilities, where weak interaction effects come into play. This could provide valuable opportunities for further investigating twist-3 effects within the hadronization process.
      
  \item  From Fig. \ref{fig:cosy} we notice that $\langle \cos\phi \rangle $ exhibits periodic characteristics.  It can be explained by considering the definition of fraction $y$,  $y=p_h\cdot l_1/p_h\cdot q \sim (1-\cos\theta)/2$, $\theta$ is the angle between the three-momenta of the electron ($\vec l_1$) and the produced hadron ($\vec p_h$). This result implies that the behavior of the asymmetry is symmetric relative to the center-of-mass as long as summing over all the quark flavors, regardless of whether the hadron is produced in the electron-beam hemisphere or the positron-beam hemisphere.
      
   \item Figure \ref{fig:cosz} shows that asymmetry  $\langle \cos\phi \rangle $ is not sensitive to fraction $z$. It might imply that asymmetry  $\langle \cos\phi \rangle $ does not depend on the FF, in other words, asymmetry  $\langle \cos\phi \rangle $  does not depend on the types of hadrons produced in the fragmentation. To verify this conclusion, we performed the numerical estimates by using the $K^+$ FFs and found the same result. However, it should be noted that this conclusion is not necessarily definitive, as there are still significant uncertainties in the current parametrizations of FFs. For example, disfavored quark FFs are often assumed to be identical in fits, $D^{\bar{u}\to K^+}(z)=D^{s\to K^+}(z)=D^{d\to K^+}(z)=D^{\bar{d}\to K^+}(z)$  \cite{Gao:2024nkz,Gao:2024dbv,Bertone:2017tyb}. Consequently, the FF contributions in the numerator and denominator in Eq. \eqref{f:cos} may cancel out. More researches therefore are needed for this topic.  
   
\end{itemize}

\begin{figure}
  \centering
  \includegraphics[width=0.9\linewidth]{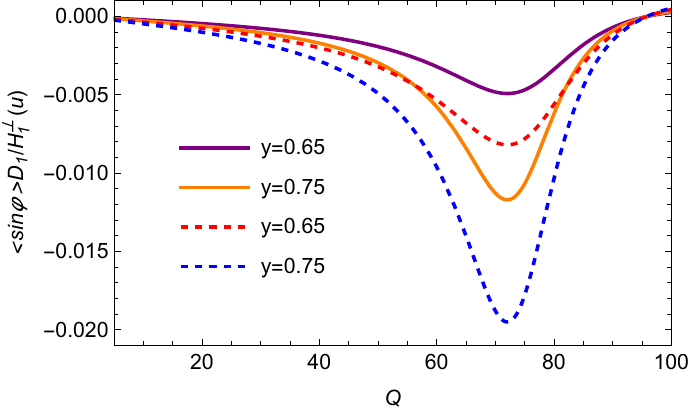}
  \caption{Numerical estimates of the chiral-odd azimuthal asymmetry $\langle \sin\varphi \rangle D_1/H^\perp_1(u)$  with respect to Q for the up quark. The fraction is taken as $z=0.3$. Solid lines represent the results for $M_J = 0.3\text{ GeV}$, while the dashed lines denote those for $M_J = 0.5\text{ GeV}$. }\label{fig:phiu}
\end{figure}

\begin{figure}
  \centering
  \includegraphics[width=0.9\linewidth]{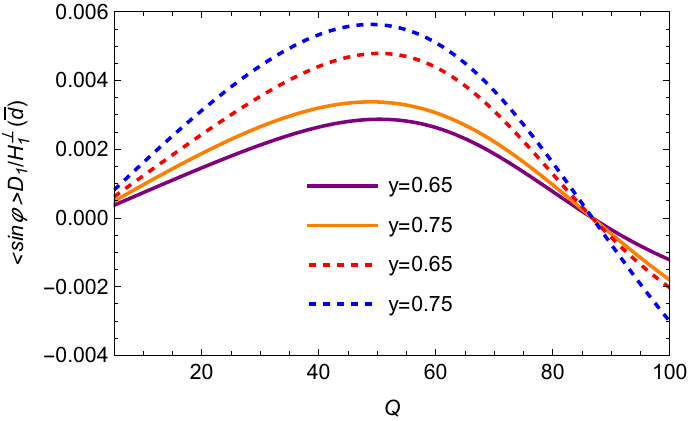}
  \caption{Numerical estimates of the chiral-odd azimuthal asymmetry $\langle \sin\varphi \rangle D_1/H^\perp_1(\bar{d})$  with respect to Q for the anti-down quark. The fraction is taken as $z=0.3$.  Solid lines represent the results for $M_J = 0.3\text{ GeV}$, while the dashed lines denote those for $M_J = 0.5\text{ GeV}$. }\label{fig:phidb}
\end{figure}

We now turn our attention to azimuthal asymmetry $\langle \sin\varphi \rangle$, which is a chiral-odd effect. Following the same methodology as described above, we define it as 
\begin{align}
   \langle \sin\varphi \rangle &=\frac{\int d\tilde{\sigma}\sin\varphi d\phi d\varphi}{\int d\tilde{\sigma} d\phi d\varphi}.
\end{align} 
Under the  Gaussian ansatz, we integrate over the transverse momentum and finally have the simplified expression,
\begin{align}
   \langle \sin\varphi \rangle &=- \frac{\Delta\sqrt{\pi}M_J}{QM}\frac{A_rt_2(y) H^\perp_1(z)}{A_r T_0(y)D_1(z)}. 
\end{align}
We notice that, unlike the asymmetry $\langle \cos\phi \rangle $ which has contributions from both EM and weak interactions, $\langle \sin\varphi \rangle $ is sensitive only to the weak interaction. Moreover,  $\langle \sin\varphi \rangle$ depends on the chiral-odd jet mass ($M_J$) or the ratio of the chiral-odd jet mass to the hadron mass ($M_J/M$), and it therefore varies for different hadron species. Providing precise numerical estimates of the azimuthal asymmetry $\langle \sin\varphi \rangle$ remains challenging, as significant uncertainties persist in the numerical results of $H_1^\perp(z)$, whether from model calculations  \cite{Matevosyan:2012ga,Bacchetta:2007wc} or parameterizations based on fits \cite{Anselmino:2007fs,Anselmino:2008jk,Anselmino:2013vqa,Anselmino:2015sxa,Kang:2015msa}. In Figs. \ref{fig:phiu} and \ref{fig:phidb}, we present the estimates of $\langle \sin\varphi \rangle D_1/H^\perp_1$ for the up quark and anti-down quark,  respectively, as they are favored quarks for $\pi^+$ production. In these estimates, the hadron mass is taken as the pion mass  ($M=139.6$ MeV) and the chiral-odd jet mass is chosen as $M_J = 0.3 \text{ GeV}$ and $0.5 \text{ GeV}$. 
Numerical estimates indicate indicates that the contributions from the up quark and anti-down quark are largely compensatory, a behavior that can likely be understood through their distinct weak coupling coefficients. We note that we did not consider the running of $M_J$, which would also introduce uncertainties, in the previous numerical estimates. 
In view of the prospective magnitude of $\langle \sin\varphi \rangle D_1/H^\perp_1$, it might serve as a helpful probe for further investigating the Collins function $H^\perp_1$. 
%From Figs. \ref{fig:phiu} and \ref{fig:phidb}, we observe that the contributions from the up quark and anti-down quark are nearly opposite, which can be attributed to their different weak coupling coefficients.  Given the sizable magnitude of $\langle \sin\varphi \rangle D_1/H^\perp_1$, it may provide a useful means to determine the Collins function $H^\perp_1$. 
 

\section{Summary} \label{sec:summary}

%Fragmentation functions serve as essential phenomenological tools for understanding hadronization. While FFs are frequently extracted from SIDIS or proton-proton collisions, electron-positron annihilation provides a particularly favorable and clean environment for studying them. In contrast to conventional approaches based on single-inclusive or semi-inclusive annihilation, the jet production one does have a simpler form of theoretical formulation and does not introduce extra uncertainties from FFs.  In particular, the invariant mass of the jet serves as a natural scale for studying chiral-odd twist-3 effects. This approach not only offers new opportunities for precise determination of FFs but also opens a path toward systematic studies of non-perturbative QCD effects within jets.

Fragmentation functions serve as essential phenomenological tools for understanding the hadronization process. While FFs are frequently extracted from SIDIS or proton-proton collisions, electron-positron annihilation offers a particularly clean and advantageous environment for their study. Compared to conventional approaches based on single-inclusive or semi-inclusive annihilation, the jet production framework can provide a more straightforward theoretical formulation, potentially avoiding certain additional uncertainties associated with FFs. Furthermore, the mass of the jet offers a natural scale for exploring chiral-odd twist-3 effects. This approach may not only facilitate a more refined determination of FFs but also open promising avenues for systematic investigations of non-perturbative QCD effects within jets.

In this paper, we present a systematic theoretical framework for calculating the differential cross-section in electron–positron annihilation to investigate twist-3 effects arising from fully inclusive jet correlators. It should be noted that we have not considered polarization effects in this calculation. For polarization effects in annihilation process, discussions can be found in Refs. \cite{Accardi:2022oog,USBelleIIGroup:2022qro}. Our calculations incorporate both electromagnetic and weak interactions. Based on the resulting differential cross-section, we explore how leading-twist FFs might be accessed through the analysis of twist-3 effects. To illustrate this, we introduce two twist-3 azimuthal asymmetries and provide the corresponding numerical estimates. Our numerical results suggest that twist-3 contributions are not as severely suppressed in the high-energy region as traditionally expected, largely owing to weak interaction effects. This indication highlights future high-energy electron–positron colliders as potentially promising platforms for searching for twist-3 effects. These results may offer valuable perspectives on the role of higher-twist dynamics in hadronization and further illustrate the feasibility of accessing leading-twist FFs through twist-3 effects.

%In this paper, we present a systematic theoretical framework for calculating the differential cross-section in electron–positron annihilation to study twist-3 effects arising from the fully inclusive jet correlators. It should be noted that we have not considered polarization effects in this calculation. For polarization effects in annihilation process, discussions can be found in Refs. \cite{Accardi:2022oog,USBelleIIGroup:2022qro}. Our calculations incorporate both electromagnetic and weak interactions.  From the differential cross-section, we notice that leading-twist FFs can be extracted by analyzing twist-3 effects. We therefore introduce two twist-3 azimuthal asymmetries and provided corresponding numerical estimates. Numerical results show that the twist-3 contributions are not significantly suppressed at high-energy region due to the weak interaction. This suggests that future high-energy electron–positron colliders offer a promising environment for observing twist-3 effects. These findings open new opportunities for investigating the role of twist-3 effects in hadronization and demonstrate the potential to extract leading-twist FFs by exploiting higher-twist effects.

\section*{Acknowledgements}
This work was supported by the National Natural Science Foundation of China (Grant No. 12405103) and the Youth Innovation Technology Project of Higher School in Shandong Province (2023KJ146).

\begin{appendix}

\section{Coordinate vectors $(t, v)$} \label{app:vector}

We first define the time-like vector $t^\mu$ and the space-like vector $v^\mu$ as
\begin{align}
   & t^\mu=\frac{q^\mu}{Q},  &&v^\mu=\frac{2p_h^\mu}{zQ}-t^\mu. \label{f:timespace}
\end{align}
These vectors satisfy the relations $t^2=-v^2=1$, and $t\cdot v=0$.  By substituting Eqs. \eqref{f:ph} and \eqref{f:q} into the above definitions, we obtain the explicit expressions, 
\begin{align}
  &  t^\mu=\frac{1}{\sqrt{2}}(\bar{n}^\mu+n^\mu) +\frac{q_T^\mu}{Q},  \label{fa:t} \\
  &  v^\mu=\frac{1}{\sqrt{2}}(\bar{n}^\mu-n^\mu) -\frac{q_T^\mu}{Q}.   \label{fa:v}
\end{align}
Using these vectors, we can construct the perpendicular tensors
\begin{align}
  & g_\perp^{\mu\nu}=g^{\mu\nu}-t^\mu t^\nu - v^\mu v^\nu,   \label{fa:g} \\
  & \varepsilon_\perp^{\mu\nu}=-\varepsilon^{\mu\nu\rho\sigma}t_\rho v_\sigma.   \label{fa:e}
\end{align}
We can substitute Eqs. \eqref{fa:t} and \eqref{fa:v} into Eqs. \eqref{fa:g} and \eqref{fa:e} and obtain,
\begin{align}
  &g_\perp^{\mu\nu}=g_T^{\mu\nu}-\frac{\sqrt{2}(\bar{n}^\mu q_T^\nu+\bar{n}^\nu q_T^\mu)}{Q}, \\
  &\varepsilon_\perp^{\mu\nu}=\varepsilon_T^{\mu\nu}+\frac{\sqrt{2}}{Q}\varepsilon^{\mu\nu\alpha\beta} \bar{n}_\alpha q_{T\beta}. 
\end{align}
In the gauge boson rest frame, $q_T=0$, we have $g_\perp^{\mu\nu}=g_T^{\mu\nu}$ and $\varepsilon_\perp^{\mu\nu}=\varepsilon_T^{\mu\nu}$. These tensors provide an alternative basis for decomposing transverse components.

\section{Coefficient functions} \label{app:func}

We begin by introducing the Wightman function
\begin{align}
  S^+_{\alpha\beta}(x-y)=\langle 0|\psi_\alpha(x)\bar{\psi}_\beta(y)|0\rangle.  
\end{align}
By inserting a complete set of states $|n\rangle$ that satisfy $\hat{P}^\mu |n\rangle= p_n^\mu |n\rangle$and applying translation invariance, the function can be rewritten as
\begin{align}
  S^+_{\alpha\beta}(x-y)=\sum_n e^{-ip_n(x-y)}\langle 0|\psi_\alpha(0)|n\rangle\langle n|\bar{\psi}_\beta(0)|0\rangle.  
\end{align}
We introduce the spectral amplitude $\rho_{\alpha\beta}(p)$ through 
\begin{align}
  \sum_n\delta^4(p-p_n)\langle 0|\psi_\alpha(0)|n\rangle\langle n|\bar{\psi}_\beta(0)|0\rangle=(2\pi)^{-3}\theta(p^0)\rho_{\alpha\beta}(p),
\end{align}
which allows us to express the Wightman function in the spectral representation,
\begin{align}
  S^+_{\alpha\beta}(x-y)=\int \frac{d^4p}{(2\pi)^3}\theta(p^0) \rho_{\alpha\beta}(p)e^{-ip(x-y)}.  
\end{align}
$\rho_{\alpha\beta}(p)$ is a scalar function only of $p$, it depends only on $p^2$, $\rho_{\alpha\beta}(p)\to \rho_{\alpha\beta}(p^2)$. Under parity constraint, we expand it as 
\begin{align}
  \rho_{\alpha\beta}(p^2)=\left[\rho_1(p^2) \slashed p +\rho_2(p^2)\right]_{\alpha\beta}.
\end{align}
By  introducing the integral over the mass spectrum, 
\begin{align}
  \rho(p^2)=\int d\mu^2\delta(p^2-\mu^2)\rho(\mu^2), 
\end{align}
we can rewrite the Wightman function $ S^+_{\alpha\beta}(x-y)$ as 
\begin{align}
   S^+_{\alpha\beta}(x-y)&=\int d\mu^2\int \frac{d^4p}{(2\pi)^3}\theta(p^0)\delta(p^2-\mu^2)\nonumber \\
   &\times\left[\rho_1(p^2) \slashed p +\rho_2(p^2)\right]_{\alpha\beta}e^{-ip(x-y)}. \label{fa:sz}
\end{align}
Similarly, we have
\begin{align}
   S^-_{\alpha\beta}(x-y)&=\langle 0|\bar{\psi}_\beta(y)\psi_\alpha(x)|0\rangle \nonumber \\
   &=\int d\mu^2\int \frac{d^4p}{(2\pi)^3}\theta(p^0)\delta(p^2-\mu^2)\nonumber \\
   &\times\left[\rho_1(p^2)(-\slashed p) +\rho_2(p^2)\right]_{\alpha\beta}e^{ip(x-y)}. \label{fa:sf}
\end{align}

We next use the equal-time limit and calculate the integral for $S^+_{\alpha\beta}(x-y)$  and $S^-_{\alpha\beta}(x-y)$. 

In the equal-time limit ($x^0=y^0$), and for the positive energy $p^0=E_p$, the integral over $p^0$ reduces to
\begin{align}
  &\int dp^0 \theta(p^0)\delta(p^2-\mu^2)f(p^0)=\frac{f(E_p)}{2E_p}. \label{fa:d0}
\end{align}
Using this result, we obtain
\begin{align}
   S^+_{\alpha\beta}(x-y)&=\int d\mu^2\frac{\left[\rho_1(p^2) \gamma^0\cdot E_p +\rho_2(p^2)\right]_{\alpha\beta}}{2E_p}\delta^3(\vec{x}-\vec{y}),  \label{fa:szint} \\
   S^-_{\alpha\beta}(x-y)&=\int d\mu^2\frac{\left[\rho_1(p^2) (-\gamma^0\cdot E_p) +\rho_2(p^2)\right]_{\alpha\beta}}{-2E_p}\delta^3(\vec{x}-\vec{y}). \label{fa:sfint}
\end{align}
Therefore, the commutator,
\begin{align}
   S_{\alpha\beta}(x-y)&=S^+_{\alpha\beta}(x-y) + S^-_{\alpha\beta}(x-y) \nonumber \\
   &=\int d\mu^2\rho_1(p^2) (\gamma^0)_{\alpha\beta} \delta^3(\vec{x}-\vec{y}) \label{fa:sint}
\end{align}
Applying the equal-time anticommutation relation
\begin{align}
  \left\{ \psi_\alpha(x),  \psi_\beta^\dag (y)\right\}=\delta_{\alpha\beta} \delta^3(\vec{x}-\vec{y}), 
\end{align}
we obtain 
\begin{align}
   S_{\alpha\beta}(x-y)&=(\gamma^0)_{\alpha\beta} \delta^3(\vec{x}-\vec{y}). \label{fa:santi}
\end{align}
From Eqs. \eqref{fa:sint} and \eqref{fa:santi}, we obtain the Kallen-Lehmann sum rule,
\begin{align}
   \int d\mu^2\rho_1(p^2)=1. \label{fa:sumrule}
\end{align}

For the jet correlator, defined as
\begin{align}
  J(p)=\sum_n\delta^4(p-p_n)\langle 0|\psi(0)|n\rangle\langle n|\bar{\psi}(0)|0\rangle,
\end{align}
the spectral representation gives
\begin{align}
  J(p)=\int d\sigma^2\left[J_1(\sigma^2) \slashed p + J_2(\sigma^2) \sigma \right]\delta(p^2-\sigma^2).
\end{align}
We integral over the light-cone component $p^+$ and obtain
\begin{align}
  J(p^-, p_T)&=\int dp^+  J(p) =\int \frac{dp^2}{2p^-} J(p) \nonumber \\
  &=\int \frac{d\sigma^2}{2p^-}\left[J_1(\sigma^2) \slashed p + J_2(\sigma^2) \sigma \right]. \label{fa:jpsigma}
\end{align}
We can also decompose $J(p)$ in terms of $\Gamma$-matrices, 
\begin{align}
  J(p^-, p_T)&=\frac{\Lambda}{2p^-} B + \frac{1}{2p^-}A \slashed p, \label{fa:jpab}
\end{align}
where $\Lambda$ is the power-counting scale. From Eqs. \eqref{f:jetfunction}, \eqref{fa:jpsigma} and \eqref{fa:jpab}, we obtain  
\begin{align}
  & \epsilon =1,  && \alpha_1 =\alpha^\perp= 1 \label{f:ea}
\end{align}
by using Eq. \eqref{fa:sumrule} and $\int d\sigma^2 J_2(\sigma^2) \sigma =M_J$ \cite{Accardi:2008ne}. Note that $A=\int d\sigma^2 J_1(\sigma^2)=1$ and $B=M_J/\Lambda$. We also note that there relations were also obtained in Ref. \cite{Accardi:2020iqn}.

\end{appendix}

%\bibliographystyle{unsrt} % 选择BibTeX样式，例如plain, apalike, unsrt等
%\bibliography{references} % references是你的bib文件名，不带.bib扩展名

\end{document}